\documentclass[prb,groupedaddress,twocolumn,floatfix]{revtex4}
\usepackage{epsfig}
\usepackage{graphicx}
\usepackage{epstopdf}
\usepackage{color}
\usepackage{amssymb,amsmath}

\begin{document}
\title{Look-alike Landau levels in locally biased twisted bilayer graphene}
\author{Tomasz Chwiej}
\email{chwiej@fis.agh.edu.pl}
\affiliation{AGH University of Science and Technology, al. A. Mickiewicza 30, 30-059 Cracow,
Poland}
\begin{abstract}
\noindent
The large lattice constant of Moire superlattice formed in twisted bilayer graphene for small twist enables 
observing the Landau levels splitting into Hofstadter butterflies in energy spectra for moderate magnetic field. 
This is expected for generic system under homogeneous bias conditions but its robustness against spatial potential 
fluctuations is left  open question. We study the energy structure of twisted bilayer system in dependence of 
both, the homogeneous magnetic field and the bias voltage applied exclusively in its central part. Although 
the translational symmetry is broken, the energy states mainly localized outside the central region may still 
condense on Landau levels and these would split revealing self-similarity feature. Moreover, besides the generic branch 
of energy states with zero-mode  Landau level at charge neutrality point, when both layers are biased with the same 
voltage, the second look-alike energy branch shifted upwards can be developed by states largely localized in central 
region. Otherwise, for counter-biasing of layers, only generic branch exists but with lowest Landau layers flanked by 
either, hole-like and electron-like states localized at the top or at the bottom layer of central biased part of 
twisted bilayer system.

\end{abstract}
\maketitle

\section{Introduction}
Twisted bilayer graphene belong to class of layered van der Waals materials\cite{vdW} and it is prepared by placing 
one graphene layer onto another, by means of CVD, pick-up or MBE 
techniques,\cite{orb_mag_exp_3,fabrication1,latt_relax_1,moire_lattice_exp,helical_exp_1} that the crystallographic axes 
in both layers differ by small twist angle $\theta$. Honeycomb graphene lattice is composed of two 
interpenetrating triangle A and B sublattices of carbon atoms and these, due to the twist angle, 
periodically move closer to and away from their counterparts in second layer leading to formation of Moire 
lattice. That introduces new length scale $L_{m}=a_{0}/2/\sin(\theta/2)$  ($a_{0}=0.246\textrm{ nm}$ is graphene 
lattice constant) defining thus spatial periodicity of interlayer, or more precisely, AA and AB/BA intersublattice 
couplings. Consequently, the energy structure is renormalized developing Moire bands irrespective 
commensurability of Moire lattice occurs or not.\cite{mele1,mele2,bistritzer2,lopes1,lopes2}
As show theoretical and experimental works, strong local 
interlayer coupling makes the particles wave functions to be largely localized at AA 
sites\cite{stm_aa_1,yan,vh1,trambly} especially for the set of magic twist angles when Fermi velocity 
is strongly suppressed due to flattening of low energy bands.\cite{bistritzer2,lopes1,lopes2,magic_origin} 
Spatial localization of Dirac particles may be further enhanced in TBLG for small twist angles $\theta<<1^{0}$ when 
atomic lattice reconstruction in both layers spontaneously occurs\cite{latt_relax_1} forming topologically 
protected one-dimensional transport channels along the edge of adjacent AB and BA 
triangles,\cite{helical_exp_1,helical_exp_2} the regions of strong pseudomagnetic field.\cite{pseudomagnetic} 
Flattening of low energy bands decreases their bandwidths making single particle kinetic energy comparable 
with electron-electron interactions in TBLG. That gives rise to spectacular correlations effects observed 
experimentally in recent years such as unconventional superconductivity,\cite{tblg_sc_1,tblg_sc_2} correlated insulating 
phases,\cite{corr_phase_1,corr_phase_2} large orbital magnetism\cite{orb_mag_exp_1,orb_mag_exp_2,orb_mag_exp_3} and
anomalous Hall ferromagnetism.\cite{anom_hall_1,anom_hall_2} 
      
Since TBLG enables engineering single particle properties, what results directly from dependence of Moire energy bands 
on twist angle, its remarkable features can be fine tuned giving rise to twisttronics.\cite{twistronics} Possibility 
of changing the Moire 
lattice constant in wide range allows to observe experimentally the Hofstadter butterflies.\cite{hofstadter} The 
fractal pattern in energy spectrum of TBLG is evoked for moderate magnetic fields when the magnetic flux 
$\phi_{M}=BS_{M}$ for single Moire cell extending over few tens of nanometers is of the same order as magnetic flux 
quantum $\phi_{0}$, unattainable condition for conventional atomic lattices. Theoretical works of Bistritzer and 
MacDonald\cite{hofstadter2} for infinite TBLG as well as of Wang et al.\cite{wang_fractal} for spatially limited TBLG 
system predicted Landau levels splitting to form fractal pattern for magnetic field range depending explicitly on twist 
angle. This effect was later observed experimentally.\cite{moire_lattice_exp,hofstadter_exp_2}

In present work we analyze formation of Landau levels in TBLG electrostatically biased in its center.
Our considerations correspond to strongly coupled twisted bilayer system that is formed for low twist 
$\theta=0.75^{0}-2^{0}$. It is known that the variations of barrier's height combined with various energies 
of incident electron in TBLG may induce not only oscillations in transmission through the barrier but even lead to its 
complete suppressing.\cite{tblg_tunneling} Thus, the combined effect of electrostatic and magnetic deflection imposed on
trajectory of Dirac particle shall substantially influence on Landau levels energy spectra. 
Our results show doubling of the number of states in energy spectra. The second look-alike Landau levels energy branch 
is shifted by about of bias voltage applied to layers provided that polarization of both layers is identical.
Otherwise, when layers are counter-biased, doubling of energy states does not occur. Instead, we observe two flanking 
states detached from the lowest Landau levels, these have electron-like or hole-like nature. 
Contrary to ordinary Landau levels, these states are localized within biased region and therefore their 
energies linearly depend on amplitude of applied voltage. For identical biasing of layers there is only one 
flanking state with constant slope depending on a sign of the bias voltage $\partial E/\partial V_{bias}\sim 
sign(V_{bias})$.

The paper is organized in the following way, in Sec.\ref{Sec:model} we briefly describe the theoretical model and 
applied numerical method, results are presented Sec.\ref{Sec:res} along with discussion, conclusions are given in
Sec.\ref{Sec:con}.

%%%%%%%%%%%%%%%%%%%%%%%%%%%%%%%%%%%%%%%%%%%%%%%%%%%%%%%%%%%%%%%%%%%%%%%%%%%%%%%%
\section{Theoretical model}
\label{Sec:model}
We consider finite size TBLG system within circular region for $r\le r_{max}=150\textrm{ nm}$. The potential 
bias is applied in its center with formula
\begin{equation}
V_{\alpha}(r)=\frac{V_{\alpha}}{1+\exp\left(\frac{r-R_{c}}{\sigma}\right)},
\label{Eq:pot}
\end{equation}
and defines the electrostatic cavity of radius $R_{c}=80\textrm{ nm}$ with respect to rest part of TBLG system,
$\alpha=t,b$ labels top and bottom layer, respectively. Two cases are analyzed in detail, first, an identical bias 
potential is 
applied to layers $V_{t}=V_{b}$ while in second case these are counter-polarized $V_{t}=-V_{b}$.
The edge of cavity is smooth and its effective width $2\sigma=10\textrm{ nm}$ is comparable with Moire length scale 
$L_{m}$ considered in work. 
That ensures optimal conditions for trajectory deflection\cite{tblg_tunneling} and separation 
of states largely localized in cetral region from the outer ones. To mimic open boundaries for 
$r>r_{out}$ and to minimize its influence on electronic states localized in central part we add complex absorbing 
potential\cite{varga,feldman,calogero} ($V_{cap}$) near the boundary $r>r_{cap}=130\textrm{ nm}$ 
\begin{equation}
 V_{cap}(r)=i\,\ln^{2}\left(1-\frac{r-r_{cap}}{r_{max}-r_{cap}} \right)   
\end{equation}
where $i=\sqrt{-1}$.
Large dimensions of considered TBLG system guarantees minimal influence of edge states on electronic 
spectrum.\cite{andelkovic}
Our main aim is to show how the Landau levels are formed in TBLG when both graphene layers are locally 
biased. In calculations we use the continuum model of TBLG which gives reasonable results for low energy 
states. The advantage of this method results from the fact that it allows us to simulate TBLG systems of larger 
dimensions than e.g. tight-binding method. On the other hand, continuum model requires much attention paid to 
spatial symmetries that has to be taken into account which issue is largely avoid when atomistic calculations are made. 
Here we use TBLG Hamiltonian proposed by Bistritzer and MacDonald\cite{bistritzer2} 
\begin{equation}
 \hat{H}_{TBLG}=\left[ 
 \begin{array}{cc}
\hat{H}_{t}(-\theta/2) & \hat{W}\\
 \hat{W}^{\dagger}& \hat{H}_{b}(\theta/2)
 \end{array}
 \right]
 \label{Eq:htblg}
\end{equation}
where $\hat{H}_{t/b}$ is single layer Hamiltonian
\begin{equation}
 \hat{H}_{\alpha}(\pm\theta/2)=\left[
 \begin{array}{cc}
  U_{\alpha} & v_{F} e^{\pm i\theta/2}\hat{\Pi}^{\dagger}\\
  v_{F} e^{\mp i\theta/2}\hat{\Pi}& U_{\alpha}
 \end{array}
 \right]
 \label{Eq:hslg}
\end{equation}
twisted by $\pm\theta/2$ with respect to x axis, $U_{\alpha}=V_{\alpha}+V_{cap}$, $v_{F}\approx10^{6}\textrm{ m/s}$ is 
Fermi velocity in graphene and $\hat{\Pi}=\hat{\pi}_{x}+i \hat{\pi}_{y}$, 
$\hat{\pmb{\pi}}=\hat{\pmb{p}}+e\pmb{A}$ where $\hat{\pmb{p}}$ is 
momentum operator. The vector potential is taken in symmetric form $\pmb{A}=B[-y,x,0]$ which leaves rotational 
symmetry of $\hat{H}_{\alpha}(\pm\theta/2)$ unchanged. 
The intersublattice coupling matrix elements are defined using identity matrix $\sigma_{0}$ and 
Pauli matrices $\sigma_{x},\sigma_{y},\sigma_{z}$ as 
$W=w_{0}\sum_{j=1}^{3}\tau_{j}e^{i\pmb{Q}_{j}\pmb{r}}$ where 
$\tau_{1}=\sigma_{0}+\sigma_{x}$, $\tau_{2}=e^{i3\pi/2} e^{i\pi/3\sigma_{z}}\tau_{1}e^{-i\pi/3\sigma_{z}} $ and 
$\tau_{3}=\tau_{2}^{*}$. 
The reciprocal lattice vectors are defined as 
$\pmb{Q}_{j}=K_{\theta}[\sin(\alpha_{j}),\cos(\alpha_{j})]$
for $\alpha_{j}=\pi/2+(j-1)2\pi/3$, 
$K_{\theta}=8\pi\sin\left(\theta/2\right)/(3a_{0})$ and 
$\pmb{r}=r[\sin(\varphi),\cos(\varphi)]$ is position vector.
The plane waves are expanded into a series of the Bessel function of first kind
\begin{equation}
 e^{i\pmb{Q}_{j}\pmb{r}}=\sum\limits_{n=-\infty}^{\infty} i^{n} J_{n}(Q_{j}r)e^{in(\varphi-\alpha_{j})}.
\end{equation}
This expansion in conjunction with explicit form of intersublattice hopping matrices $\tau_{i}$ gives 
the intersublattice  coupling elements $w_{\mu\nu}$ $(\mu,\,\nu=\{A,B\})$
\begin{equation}
 w_{\mu\nu}=3w_{0}\sum\limits_{k=-\infty}^{\infty}J_{3k+\eta}(Qr)e^{i(3k+\eta)\varphi}
 \label{Eq:vmn}
\end{equation}
where $\eta=+1,0,-1$ for $w_{AA}=w_{BB}$, $w_{AB}$ and $w_{BA}$, respectively. 
The strength of interlayer coupling is scaled by $w_{0}=110\textrm{ meV}$, however in unpatterned TBLG due to surface 
corrugation its value can be smaller for AA sites than for Bernal stacking regions.\cite{corrugation}
Eigenvectors of TBLG Hamiltonian given in Eq.\ref{Eq:htblg} are four-component spinors
$\Psi(\pmb{r})=\left[\psi_{A_{t}}(\pmb{r}),\psi_{B_{t}}(\pmb{r}),\psi_{A_{b}}(\pmb{r}),\psi_{B_{b}}(\pmb{r}) 
\right]^{T}$. We use finite element method to solve this eigenvalue problem. 
The interlayer coupling potentials expressed in Eq.\ref{Eq:vmn} explicitly depends on set of angular 
momenta ($l=3k+\eta$) and keeping in mind that single layer Hamiltonians have rotational symmetry it is 
reasonable to project the problem onto cylindrical coordinates. 
Then components of the four-spinor can be expressed in a basis of products of radial $\{f_{m}(r)\}$  and 
angular momentum states $\{e^{il\varphi}\}$
\begin{equation}
 \psi_{\lambda}(r,\varphi)=\sum\limits_{m,l_{\lambda}}  c_{\lambda,m,l_{\lambda}}f_{m}(r)e^{il_{\lambda}\varphi}
\label{Eq:wavefun} 
\end{equation}
where $\lambda \in\{A_{t},B_{t},A_{b},B_{b}\}$ labels particular sublattice, $m$ enumerates the radial elements 
taken as Hermite polynomials, $l_{\lambda}$ is the angular momentum and $c_{\lambda,m,l_{\lambda}}$ are linear 
expansion coefficients. With the above form of wave function the matrix elements of coupling potential take simple 
real-value form
\begin{align}
\nonumber
&\langle f_{m_{\mu}}e^{il_{\mu}\varphi}|w_{\mu\nu}|f_{m_{\nu}}e^{il_{\nu}\varphi}\rangle
=\\
\nonumber
& 6\pi (-1)^{\max(0,\,l_{\mu}-l_{\nu})}\,\delta_{l_{\mu}-l_{\nu},3k+\eta}\\
&\times \int\limits_{0}^{r_{max}}dr\,r\,f_{m_{\mu}}(r)\,f_{m_{\nu}}(r)\,J_{|l_{\mu}-l_{\nu}|}(Qr)
\label{Eq:matel}
\end{align}
where $r_{max}$ is the radius of system, $\delta_{l_{\mu}-l_{\nu},3k+\eta}$ is Kronecker's delta which 
matches the angular momentum of both, the upper and the lower layer's states, with infinite series defining the 
coupling 
potential (Eq.\ref{Eq:vmn}).  Owing to this fact, the non-zero 
elements generally constitute an infinite sequence for  $l_{\mu}-l_{\nu}=3k+\eta,\quad k=0,\pm 1,\pm 2,\ldots$. 
Actually for finite $r_{max}$ this sequence also becomes finite because the Bessel function $J_{|l_{\mu}-l_{\nu}|}$ 
in the integral (Eq.\ref{Eq:matel}) quickly tends to zero for increasing angular momentum difference 
$|l_{\mu}-l_{\nu}|$.
Finally the eigenvalue problem takes generalized form $ \pmb{H}\Psi=E\pmb{S}\Psi$
which was effectively solved by exploiting sparsity of Hamiltonian and overlap integrals matrices. 
In calculations the radial elements have equal width $\Delta r=2\textrm{ nm}$ giving 75 elements while the maximal  
angular momentum that gives rise to energy is limited to $|l_{\lambda}|\le 360$.
Artificial absorbing potential ($V_{cap}$) included in our model, on one hand allows to mimic the nanostructure of 
infinite dimensions, but on the other hand it destroys the hermiticity of TBLG Hamiltonian. The imaginary part of 
eigenenergy 
($E_{\alpha}=\varepsilon_{\alpha}-i\gamma_{\alpha}$) determines the coupling strength between the $\alpha$ 
state and the continuum part of energy spectra, its small value means weak coupling and vice versa. 

Diagonalization provides us with two sets of the right and the left  wave vectors $\{\Psi_{i}^{(r/l)}\}$ which 
are four-spinors $\Psi_{i}^{(r/l)}=\sum_{\mu}\psi_{i,\mu}^{(r/l)}\pmb{e}_{\mu}$ and $\pmb{e}_{\mu}$ are the 
4-dimensional Cartesian basis vectors. These we use to approximate the retarted Green function
in this space $G^{R}(E)=\sum_{\mu,\nu}G^{R}_{\mu,\nu}(E) \pmb{e}_{\mu}\otimes \pmb{e}_{\nu}$ with matrix elements 
$G_{\mu\nu}^{R}(E)=\sum\limits_{i}\psi_{i,\mu}^{r}(\psi^{l}_{i,\nu})^{*}/(E-E_{i})$. Having $G^{R}$ we calculate the 
matrix elements of spectral function $\rho_{\mu\nu}(\pmb{r},\pmb{r'},E)=(i/2\pi)[G^{R}-(G^{R})^{\dagger}]_{\mu\nu}$, 
density of states (DOS) $D(E)=Tr\{\rho(\pmb{r},\pmb{r'},E)\}$, local density of states (LDOS) 
$\rho(\pmb{r},E)=\sum_{\mu}\rho_{\mu\mu}(\pmb{r},\pmb{r},E)$ and, x and y components of magnetic current 
($\pmb{j}=dH/d\pmb{A}$)
$\pmb{j}(\pmb{r},E)=-qv_{F}\sum_{\mu,\nu}\rho_{\mu,\nu}(\pmb{r},\pmb{r},E)\pmb{e}_{\nu}^{T}(\tau_{0}\otimes 
\pmb{\sigma})\pmb{e}_{\mu}$. These quantities are defined for given energy $E$ which in e.g. scanning tunneling 
spectroscopy (STS) is not well defined due thermal smearing, for this reason we average considered quantity $O$ over 
energy for temperature $T=4.2\textrm{ K}$ as an integral $\langle O\rangle=\int dE'\, w(E',E,T)\, O(E')$ with 
window 
function taken as derivative of Fermi-Dirac distribution function $w(E',E,T)=-df(E',E,T)/dE'$. Window function scales 
contributions of energy states in transport measurements near given energy E.\cite{datta} 
Besides the temperature also complex absorbing potential gives rise to broadening of DOS as it introduces 
disorder at the edge of TBLG.\cite{andelkovic}

%%%%%%%%%%%%%%%%%%%%%%%%%%%%%%%%%%%%%%%%%%%%%%%%%%%%%%%%%%%%%%%%%%%%%%%%%%%%%%%%
\section{Results and discussion}
\label{Sec:res}

\begin{figure}[htbp!]
\hbox{
 	\epsfxsize=80mm     
       \epsfbox[0 0 1500 750] {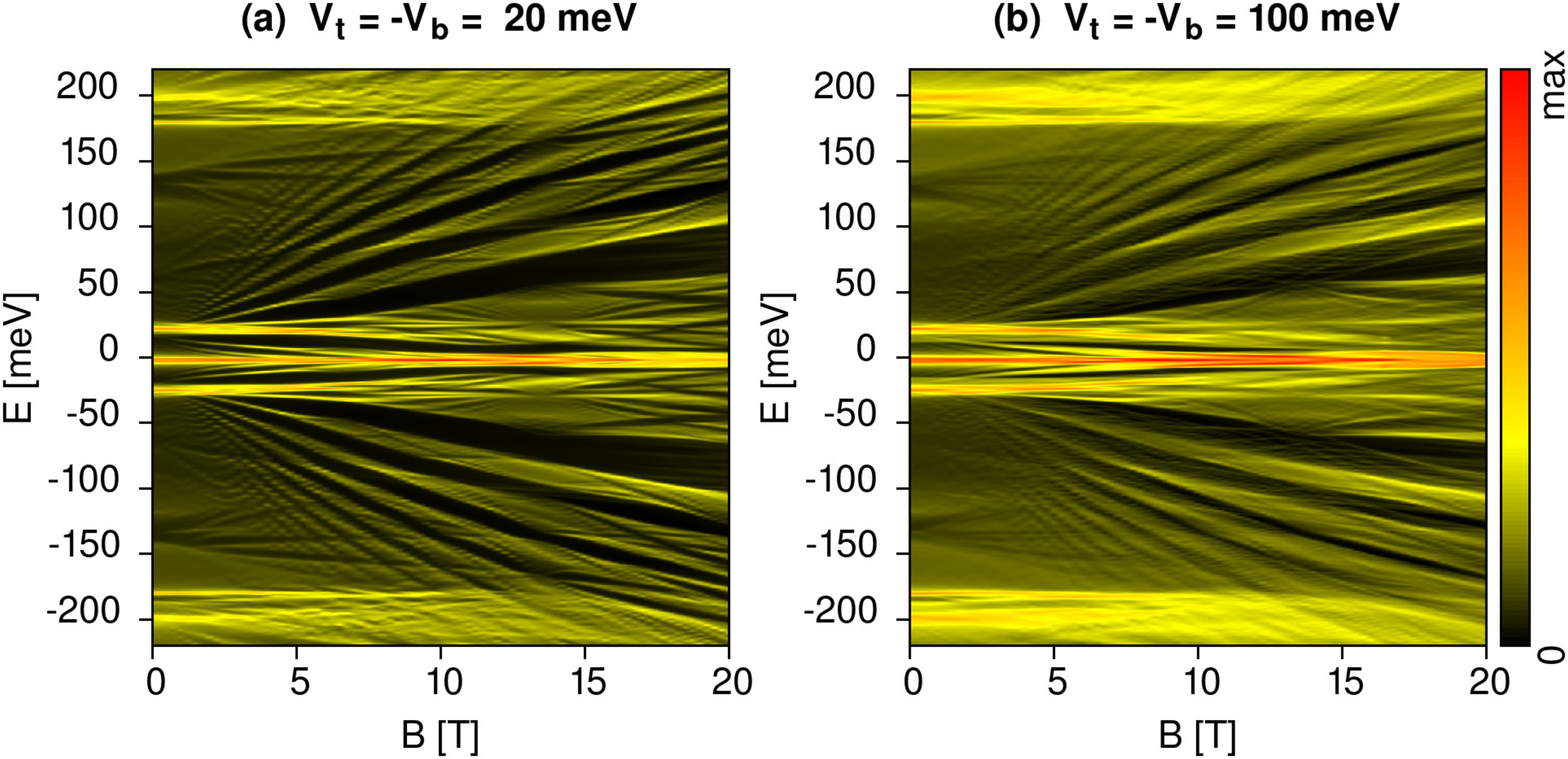}      
\hfill}
\hbox{
 	\epsfxsize=80mm     
       \epsfbox[0 0 844 236] {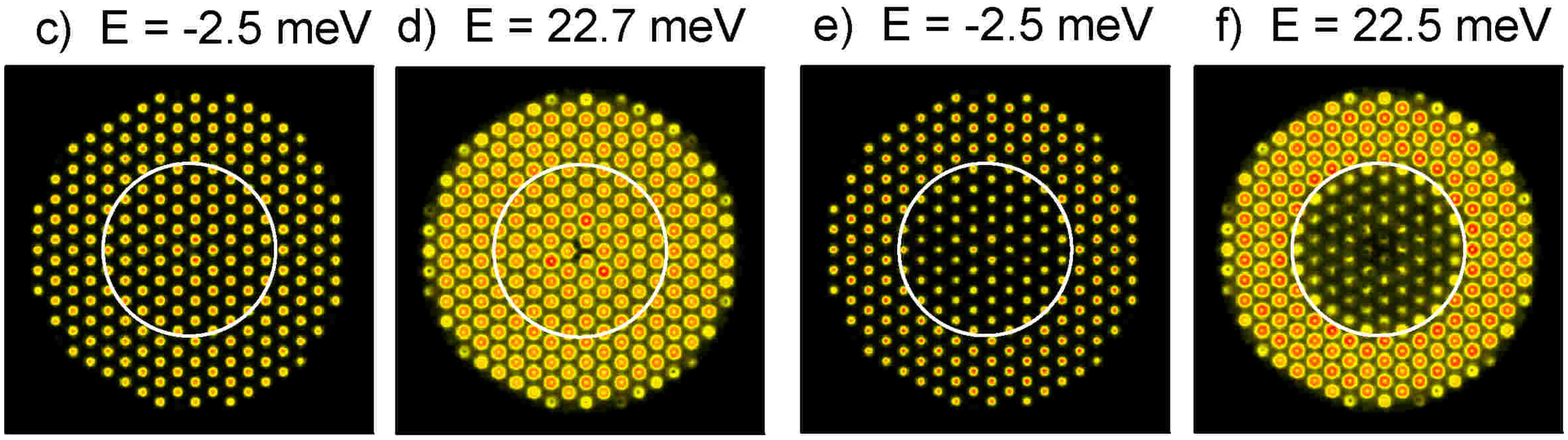}     
\hfill}
\caption{(Color online) DOS (upper row) and LDOS (lower row) for TBLG with central biased region, parameters 
used in calculations: $\theta=0.75^{0}$, $V_{t}=-V_{b}=20\,\textrm{meV}$ (a,c,d) and 
$V_{t}=-V_{b}=100\,\textrm{meV}$ (b,e,f). In (c)-(f) obtained for $B=1\textrm{ T}$ energies are displayed on top while 
the thick white circle marks approximately an edge of cavity ($R_{c}=80\textrm{ nm}$).}
\label{Fig:t075}
\end{figure}

Results presented in this section were obtained for local biasing TBLG within a circle area of radius $R_{c}=80\, nm$ 
with edge smoothed by factor $\sigma_{c}=5\, nm$ which is constant for three 
considered twist angles $\theta=0.75^{0},\, 1.05^{0}\,\textrm{and}\, 2^{0}$.
These give set of Moire lattice constants $L_{m}=18.8,\,13.4,\textrm{and}\, 7.05\, \textrm{nm}$.

\begin{figure}[htbp!]
\hbox{
 	\epsfxsize=80mm     
        \epsfbox[0 0 1500 1500] {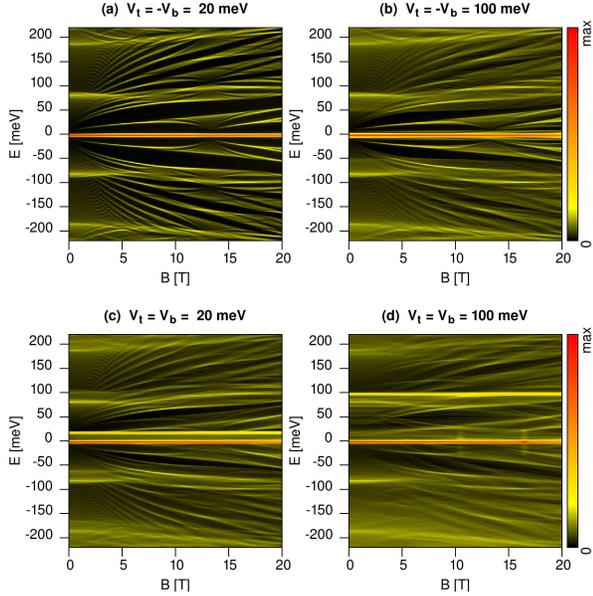}     
\hfill}
\caption{(Color online) TBLG energy spectra for $\theta=1.05^{0}$ with counter-polarization of layers
$V_{t}=-V_{b}$ (first row) and the same polarization $V_{t}=V_{b}$ (second row) of layers in central region.}
\label{Fig:teta1}
\end{figure}

DOS calculated for $\theta=0.75^{0}$ is shown in Fig.\ref{Fig:t075}. Even though top and bottom layers in central part  
are biased asymmetrically [Fig.\ref{Fig:t075}(a)], the particle-hole symmetry is slightly broken due to interlayer 
coupling $w_{AA}$.\cite{magic_origin} For this reason the  charge neutrality point (CNP) localizes at $E\approx 
-2\textrm{ meV}$ where two van Hove singularity peaks merge.\cite{trambly,lopes2,vh1,vhs_exp} Irrespective of magnetic 
field variations CNP does not change its energy and forms zero-mode Landau level (ZMLL).\cite{zmll_1}
It is flanked  by two satellites shifted by $\pm 24\textrm{ meV}$ which positions are robust against changes of 
amplitude of polarization potential [cf. Figs.\ref{Fig:t075}(a) and (b)]  but are splitted for magnetic field 
$B>5\,T$. Note however that ZMLL splits for\cite{pilkyung} $B>B_{c}\approx 3.3\,\theta^2=1.9\,\textrm{T}$ as for generic 
TBLG. At moderate magnetic field strength, the magnetic effects dominate the kinetic energy of Dirac particles and 
Landau levels becomes easily recognizable due to their characteristic B dependence $E_{n}=E_{CNP}+sgn(n)\sqrt{2e\hbar 
v_{f}^{2}|n|B}$ for integer n. Due to relatively large Moire lattice constant ($L_{m}=18.8\textrm{ nm}$) the magnetic 
flux piercing Moire unit cell $\phi_{m}$ and quantum of magnetic flux $\phi_{0}$ are comparable. For such conditions, by 
virtue of Hofstadter theory,\cite{hofstadter} each single energy band must split what manifests in self-similarity 
of resultant energy spectrum.
Despite occurrence of distinct wide LLs fans in Figs.\ref{Fig:t075}(a) and \ref{Fig:t075}(b) for $\theta=0.75^{0}$,
formation of fractal pattern in energy spectra (Moire butterflies\cite{hofstadter2,moire_lattice_exp}) is hardly 
recognizable.  This feature becomes apparent for lager twist angle considered here, namely $\theta=1.05^{0}$, what we 
notice in Fig.\ref{Fig:teta1}(a) around $B=14\textrm{T}$. Direct comparison of four cases presented in this figure 
reveals that low opposite biasing of layers is preferable to other conditions [cf. Fig.\ref{Fig:teta1}(a) and  
Figs.\ref{Fig:teta1}(b)-(d)] which largely suppress this subtle effect.

LDOS calculated for $\theta=0.75^{0}$ and $B=1\textrm{ T}$ presented in Figs.\ref{Fig:t075}(c)-(d) for two VHS peaks 
shows that for low opposite bias ($V_{t}=-V_{b}=20\textrm{ meV}$) it is composed of AA centered 
spatially-separated point-like as well as small ring-like density grains which spreads over generic TBLG and biased 
region. That partly results from applying potential difference to layers since it opens a gap in 
Bernal stacking AB/BA sites.\cite{ab_gap} Then at contacts of these regions, on a line connecting closest AA sites, 
one-dimensional transport channels are formed which are topologically protected for small twist 
angles($\theta<0.5^{0}$).\cite{helical1,helical_displacement,helical_displacement_2,helical_exp_1,helical_exp_2}

\begin{figure}[htbp!]
\hbox{
 	\epsfxsize=70mm     
        \epsfbox[0 0 1467 1084] {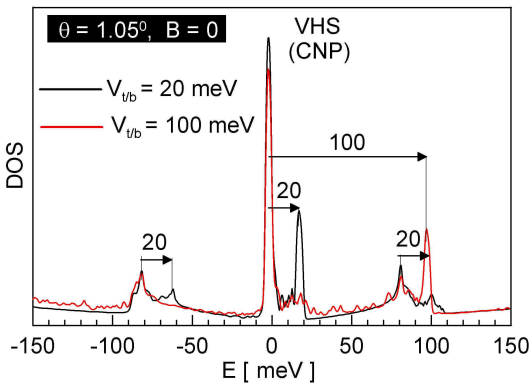}     
\hfill}
\caption{(Color online) DOS for TBLG with twist $\theta=1.05^{0}$, $B=0$, and two cavity's potentials 
$V_{t/b}=20\textrm{ meV}$ (black) and $V_{t/b}=100\textrm{ meV}$ (red). Arrows with numbers show direction and 
approximate shift of dubbed states.}
\label{Fig:dosteta1}
\end{figure}

\begin{figure}[htbp!]
\hbox{
 	\epsfxsize=70mm     
        \epsfbox[0 0 795 1079] {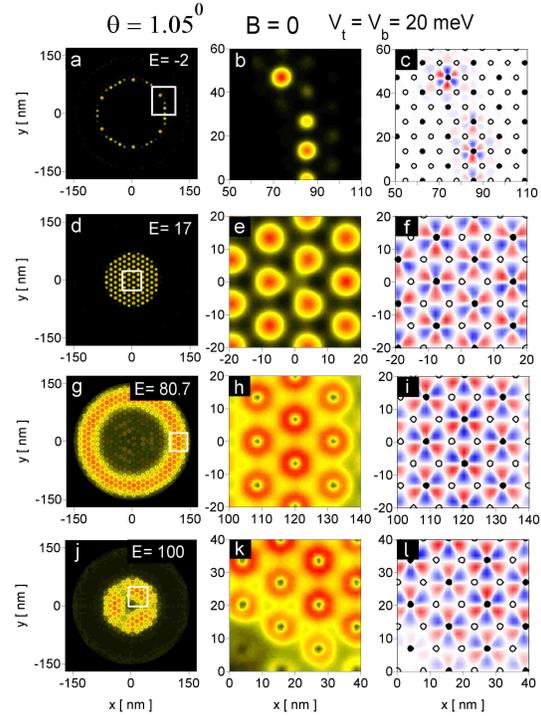}     
\hfill}
\caption{(Color online) LDOS (first and second columns) and vorticity of current (third column) for $\theta=1.05^{0}$, 
bias potential 
$V_{t}=V_{b}=20\textrm{ meV}$ and $B=0$. Energies displayed in first column correspond to DOS peaks presented in 
Fig.\ref{Fig:dosteta1} (black). Full (empty) dots in vorticity maps show positions of AA (AB/BA) sites.}
\label{Fig:ldos1b0pp20}
\end{figure}

As already mentioned DOS for first magic angle ($\theta=1.05^{0}$) displayed in Figs.\ref{Fig:teta1}(a) and (b) 
show more distinct manifestation of fractal structure of LLs which are shifted towards slightly stronger magnetic 
fields. 
Position of ZMLL do not change but VHS satellites move away by $\Delta E_{1}\approx \pm 81\textrm{ meV}$ and 
again their positions are independent of amplitude of applied bias provided that layers in center of TBLG are 
counter-polarized [cf. Figs.\ref{Fig:teta1}(a) and (b)]. That picture will change if both layers are identically 
polarized what show Figs. \ref{Fig:teta1}(c) and \ref{Fig:teta1}(d). Since biasing of graphene layers in central region 
is the same the energy structure captured within it is pushed up by about  $\Delta E\approx V_{t/b}$. The most striking 
evidence of doubling the number of energy levels is occurrence of second ZMLL which stays insensitive to magnetic 
field. Moreover in Fig.\ref{Fig:teta1}(c) we see that both DOS satellites are also replicated as there occur two, but 
less intensive, DOS peaks shifted upwards by about $20\textrm{ meV}$. For stronger central biasing 
($V_{t}=V_{b}=100\textrm{ meV}$) this shift could be hardly resolved, besides strong look-alike ZMLL, but undeniably 
must exist since crossings of LLs are visible even in moderate magnetic field $B=5-10\textrm{ T}$ [see 
Fig.\ref{Fig:teta1}(d)].

\begin{figure}[htbp!]
\hbox{
 	\epsfxsize=40mm     
        \epsfbox[0 0 278 265] {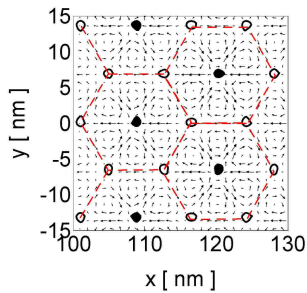}     
\hfill}
\caption{(Color online) Vector plot of current for the same parameters as in Fig.\ref{Fig:ldos1b0pp20}(g) 
($E=80.7\textrm{ meV}$).}
\label{Fig:curb0pp20}
\end{figure}

Doubling of energy structure for identical biasing of layers becomes most distinct for vanishing magnetic field.
Figure \ref{Fig:dosteta1} shows two DOS spectra for $B=0$ which do not change much for $B<5\textrm{ T}$.
We see that central (CNP) and flanking VHS peaks are duplicated and their copies, few times smaller, are shifted 
upwards in energy in accordance with bias potential applied to both layers $V_{t}=V_{b}$.
LDOS maps displayed in Fig.\ref{Fig:ldos1b0pp20} show that CNP and its neighbouring VHS are pushed outside cavity 
(first and third row) what explains their insensitivity to variations of bias potential. That differ them from their 
shifted counterparts, which as expected, are largely localized in cavity [see Figs.\ref{Fig:ldos1b0pp20}(d) and 
(j)]. Although cavity has finite extensions, for $R_{c}=80\textrm{ nm}$ and $L_{m}=13.4\textrm{ nm}$ it covers 
$n_{m}\approx 4\pi R_{c}^{2}/(\sqrt{3} l_{m}^{2})\approx 257$ Moire supercells, enough to develop additional  
energy quasi-bands separated from the ones formed for unbounded and unbiased rest part of TBLG.    
Even though, the renormalized Fermi velocity is considerably suppressed in vicinity of each magic 
angle\cite{bistritzer2} enhancing thus particle localization around AA sites for low energy [see 
Figs.\ref{Fig:ldos1b0pp20}(a),(b), (d) and (e)], the tunneling on Moire lattice between AA sites is supported by 
helical current. An example of current density is shown in Fig.\ref{Fig:curb0pp20} which in considered cases is hardly 
readable, instead we will show vorticity of current $v_{j}=(\nabla\times\vec{j})_{z}$ which nodal surface separates 
countercirculating currents. Pattern of current vorticity [last column in Fig.\ref{Fig:ldos1b0pp20}], similar for all AA 
sites with respect to local 
variations in intensity, is composed of triangle-shaped three current vortices and three antivortices connected at AA 
sites. Due to bending of particle's trajectory within each current vortex only its edge part can couple to 
neighbouring counter-oriented vortex, at midway between AA sites, by bending the wave vector drawing 
thus simple or reflected elongated S-like path. Surprisingly, even though  density rings centered at AA sites are 
weakly connected with straight bridges [Figs.\ref{Fig:ldos1b0pp20}(e) and (h)], the pattern of current vorticity does 
not change. This specific orbital antiferromagnetic property of Moire supercells is valid only if interactions are 
weak, otherwise, as shown in Ref.[\cite{electric_magnetism}], applying large interlayer bias potential may trigger 
transition from the lattice antiferromagnetic phase to the spiral ferromagnetic one in TBLG. Interactions can also 
enhance spin and valley polarizations triggered by variations of symmetry breaking small magnetic field giving rise to 
anomalous Hall effect.\cite{anom_hall_1,anom_hall_2} Since in considered system the interlayer biasing is local we 
expect that such 
geometry would, under properly chosen conditions, allow to create and control different magnetic phases in adjacent 
spatial regions.

\begin{figure}[htbp!]
\hbox{
 	\epsfxsize=90mm      
       \epsfbox[0 0 1500 1500] {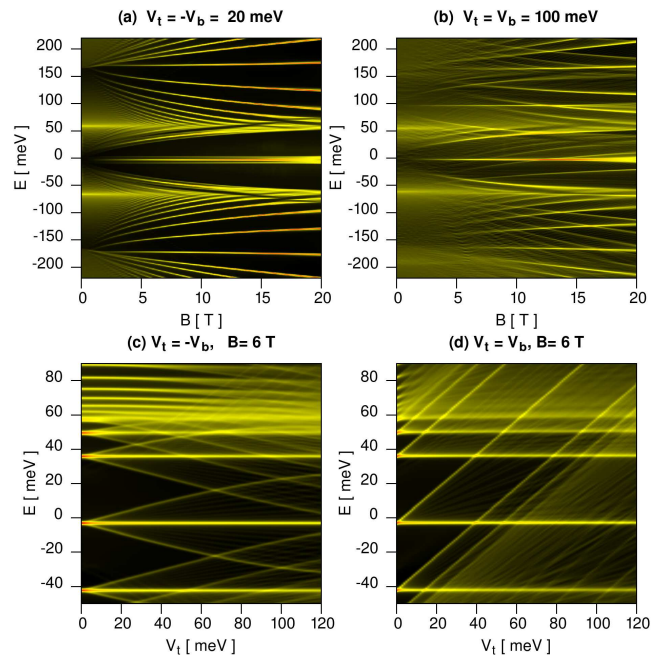}
\hfill}
\caption{(Color online) DOS for $\theta=2^{0}$, $R_{c}=80\textrm{ nm}$ in function of B for $V_{t}=-V_{b}=20\textrm{ 
meV}$ (a) and the same polarization of layers $V_{t}=V_{b}=100\textrm{ meV}$ (b), figures (c) and (d) show 
LLs' satellites for counter- and the same polarization of layers at $B=6\textrm{ T}$.}
\label{Fig:teta2}
\end{figure}

\begin{figure}[htbp!]
\hbox{
 	\epsfxsize=70mm      
       \epsfbox[0 0 795 1076] {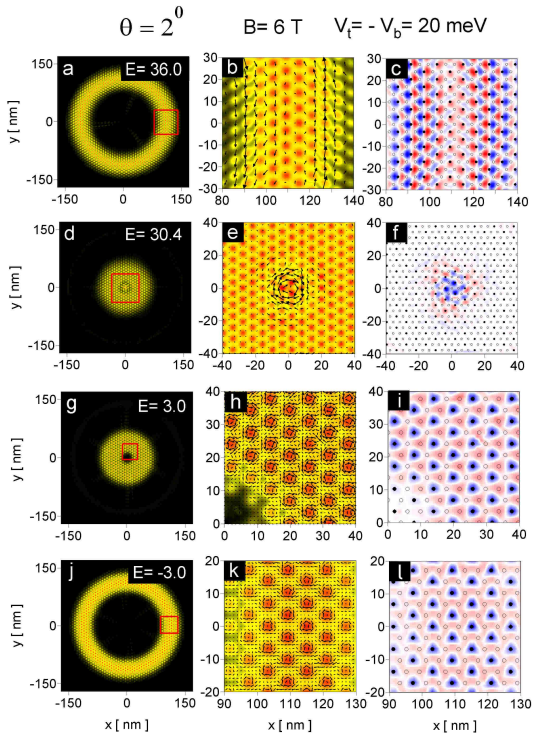}
\hfill}
\caption{(Color online) LDOS (left column), its enlarged part (red rectangle) with vector plot of current (middle 
column) and vorticity of current (right column) for zero mode LL (lowest row) and next LL (top row) and their 
satellites (middle rows) calculated for $V_{t}=-V_{b}=20\textrm{ meV}$ and $B=6\textrm{ T}$. Full (empty) dots in  maps 
of current vorticity  show positions of AA (AB/BA) sites.}
\label{Fig:ldos_t2} 
\end{figure}

Increasing further the twist angle to $\theta=2^{0}$  substantially changes the DOS evolution in magnetic field.  
Figure \ref{Fig:teta2} shows that states will start to condense on LLs even for small magnetic field ($B>2\,T$) and 
moderate energies ($E\approx 150\textrm{ meV}$). Moreover, due to shifting van Hove's singularities to $E=-66\textrm{ 
and } 59\textrm{ meV}$ 
($\Delta E_{VHS}=125\textrm{ meV}$) DOS at CNP is minimal but zero mode LL is restored for higher B.
Positions of VHS peaks in generic TBLG can be approximated\cite{vhs_exp} as $\Delta E_{VHS}=\hbar 
V_{F}\Delta K-2 t^{\theta}$ which for $\theta=2^{0}$, $t^{\theta}\approx 0.4\, t_{\perp}$ and $t_{\perp}=3\,w_{AA}$ 
gives $127\textrm{ meV}$. Although  LDOS for maxima of VHS go outside the cavity, similarly as for $\theta=1.05^{0}$, 
there was achieved satisfactory agreement. It is worth to note however that value of $t_{\perp}$  we used
is noticeably larger from originally proposed ones\cite{vhs_exp} ($t_{\perp}=240 \textrm{ and } 270\textrm{ meV}$), on 
the other hand the dependence of VHS positions on twist angle generally varies for different substrate the TBLG is put 
on.\cite{vhs_exp_3,vhs_exp_4}

We have checked that small interlayer bias ($V_{t}=V_{b}=20\textrm{ meV}$) as shown in Fig.\ref{Fig:teta2}(a) 
gives almost similar pattern as for unbiased case besides the small satellite peaks flanking lowest LL's. These 
satellites are sensitive to the variations of bias potential what we notice in Fig.\ref{Fig:teta2}(c). 
Each pair of satellites  originates from single LL state and increasing bias potential shifts positions of 
flanking states towards higher or lower energy as if most parts of these states are accumulated at upper or at lower 
layer. They may cross with other ones originating from neighbouring LLs.
Additionally, we see in Fig.\ref{Fig:teta2}(c) that the lowest LLs are insensitive to bias potential even though it 
extends over a large area in center of TBLG. Such unusual behaviour we explain by analyzing LDOS for ZMLL, first LL 
and their two satellites which are shown in Fig.\ref{Fig:ldos_t2} for $B=6\textrm{ T}$.
LDOS for ZMLL ($E=-3\textrm{ meV}$) as well as for first LL ($E=36\textrm{ meV}$) is pushed outside the cavity and due 
to magnetic deflection both form a ring-like structure. The enlarge parts of LDOS (second column) show however, that 
ZMLL density islands are localized  at AA sites with largely disconnected current loops 
circulating around each AA site.
Although, all sites AA gives non-zero net magnetization for ZMLL what we deduce from the current vorticity, these are 
surrounded by AB and BA regions with opposite vorticity leading eventually to their cancellation. For 1LL 
[Figs.\ref{Fig:ldos_t2}(b) and (c)] maxima of LDOS at AA sites becomes less distinct, here however current loops from 
neighbouring AA sites merge and consequently the density current flows in clockwise (counterclockwise) direction on 
inner (outer) side of ring-like LDOS. Current flowing through AA and AB/BA sites is only slightly locally deflected but 
its global orientation remains unchanged.      
Second and third rows in Fig.\ref{Fig:ldos_t2} shows results for LLs' satellites. Both are largely localized inside of 
central region and therefore they must be sensitive to interlayer bias potential. Despite stronger 
accumulation of these states at upper or lower layer, their properties are still remarkably influenced by interlayer 
coupling. The one which decouples from ZMLL (third row) has distinct AA island-like LDOS structure with current loops 
circulating around similarly as for ZMLL. Second satellite (second row) besides clear triangle pattern  in 
LDOS develops  current vortex in the very center of TBLG while the current almost vanishes outside.

Contrary to this case, by applying the same potential 
to both layers [see Fig.\ref{Fig:teta2} (b)] makes DOS spectra more complicated even for large magnetic field where LLs 
cross each other. In this case each LL has only one satellite shifted towards higher energy due to identical 
polarization of layers [see Fig.\ref{Fig:teta2}(d)]. These, however, are less pronounced than LLs and can 
not form such distinctive crossings as we see in Fig.\ref{Fig:teta2}(b).     
Because spatial size of considered cavity is large ($R_{c}=80\textrm{ nm}$) as compared to present Moire lattice 
constant ($L_{m}=7.05\textrm{ nm}$) that gives enough space to develop look-alike LLs 
structure shifted upwards by $V_{t}=V_{b}=100\textrm{ meV}$. We have confirmed this by conducting additional 
calculations for $V_{t}=V_{b}$ but limiting the biased region to $R=50\textrm{ nm}$ (results not show here). 
The outcomes showed only generic TBLG DOS states merging into LLs as in Fig.\ref{Fig:teta2}(a) because  central 
biased region has not enough space to develop its own energy pseudobands. The look-alike energy states we already 
observed for $\theta=1.05^0$ in Figs.\ref{Fig:teta1} (c) and (d), where second ZMLL state has emerged for 
$E=20\,\textrm{and}\,100\textrm{ meV}$ accordingly with bias potential. However, due to much attenuated DOS spectra in 
those cases we can not definitely claim if the whole low-energy spectra was duplicated or only its part.
Certainly, very recognizable peaks localized near $E\sim \pm 81\textrm{ meV}$ [see Fig. \ref{Fig:dosteta1}] have  
counterparts shifted upward in energy by $V_{t}=V_{b}=20\,\textrm{meV}$. 
Other look-alike states, if even exist for larger B, are hardly recognizable due to strongly broaden spectra.

\begin{figure}[htbp!]
\hbox{
 	\epsfxsize=90mm      
	\epsfbox[0 0 1500 750] {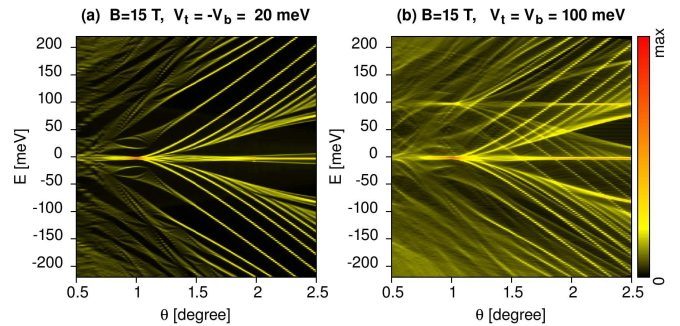}
\hfill}
\caption{(Color online) Energy spectra of TBLG for parameters $R_{c}=80\textrm{ nm}$, $B=15\textrm{ T}$, 
$V_{t}=-V_{b}= 20\textrm{ meV}$ (a) and $V_{t}=V_{b}=100\textrm{ meV}$ (b) in function of the twist angle.}
\label{Fig:eteta}
\end{figure}

Results presented so far indicate that twist angle plays crucial role in doubling the number of LL states.
To confirm this we will analyze DOS in function of twist angle, results for identical and counter-biasing of layers in 
strong magnetic field ($B=15\, T$) are presented in Fig.\ref{Fig:eteta}.    
Interestingly, in counter-biasing case [Fig.\ref{Fig:eteta}(a)] the low energy spectrum develops fractal pattern for 
$\theta\le 1.05^{0}$ as well as for larger energies until $\theta<1.5^{0}$. Increasing twist angle beyond first magic 
angle 
separates the 
ZMLL from other LLs. Due to both, counter-biasing and strong magnetic field,
the flanking satellite states are hardly visible besides the ZMLL for $\theta>2^{0}$. 
However, by applying the same bias potential to layers substantially changes energy spectra. For 
$V_{t}=V_{b}=100\textrm{ meV}$ we easily recognize in Fig.\ref{Fig:eteta}(b) look-alike second LLs branch. These states 
are shifted upwards and surprisingly reconstruct also the self-similarity 
feature of energy spectra for $\theta<1.05^{0}$.

%%%%%%%%%%%%%%%%%%%%%%%%%%%%%%%%%%%%%%%%%%%%%%%%%%%%%%%%%%%%%%%%%%%%%%%%%%%%%%%%
\section{Conclusions}
\label{Sec:con}

We used continuum model to study the Landau levels formed in twisted bilayer graphene with bias potential applied in 
its center. Although the electrons can not be confined definitely in space due to gapless 
energy structure of generic TBLG, the combined effect of magnetic and electrostatic deflection applied on their 
trajectories can enhance their momentary spatial localization and thus largely enrich resultant energy spectrum. 
Namely, we observe formation of distinct Landau levels with characteristic fractal pattern arising in moderate magnetic 
field ($B<20\textrm{T}$) for small twist angle $\theta<2^{0}$ provided that layers in central region are 
counter-biased. In this case, the lowest LLs are localized outside cavity forming a ring-like structure. When the 
same bias is applied to both layer, the energy spectrum becomes messy as it contains two branches of LLs crossing each 
other. The second branch is shifted in energy according to applied bias and is developed by states strongly localized 
within spatially limited biased region. These look-alike energy states can be formed provided that the ratio of biased 
region size and Moire lattice constant is large enough which we estimate to be at least $R_{c}/L_{m}>5\div 7$. 
For identical and counter-biasing of layers each Landau level becomes a precursor of one 
(electron-like or hole-like) and two (electron-like and hole-like) flanking states, respectively. Energies of 
these satellites change approximately linearly with bias potential since they are largely localized in center of TBLG 
system. 
Since density of states in TBLG can be sampled in STM experiments locally giving the same pattern irrespective of 
spatial position over TBLG plane,\cite{wang_fractal} we think the existence of described here look-alike states as well 
as flanking states could be verified at least for case with identical biasing of layers in TBLG nanodevice with split 
back gate. This would be possible for $\theta\sim 2^{0}$ while for smaller twist angles $\theta\approx 1.1^{0}$, as 
show STM experiments,\cite{vhs_exp_2}  occurrence of spatial distortions in Moire lattice when energy of tunneling 
electrons is tuned to VHS positions may suppress considered effects.

%%%%%%%%%%%%%%%%%%%%%%%%%%%%%%%%%%
\section*{Acknowledgements}
This work was (partially) supported by the AGH UST statutory tasks No. 
11.11.220.01/2 within subsidy of the Ministry of Science and Higher Education.
\section*{References}
\bibliography{tblg_landau_references}
\end{document}